\documentclass[twocolumn,aps,prc,superscriptaddress,floatfix]{revtex4}
\usepackage{amssymb}
\usepackage{amsmath,bm}
\usepackage[normalem]{ulem}
\usepackage[dvips]{color}
\usepackage{booktabs}
\usepackage{threeparttable}
\usepackage{graphicx}
\usepackage{CJK}
\usepackage[colorlinks,
            linkcolor=blue,
            anchorcolor=blue,
            citecolor=blue]{hyperref}
\renewcommand{\sout}{\bgroup \color[rgb]{1,0,0}\ULdepth=-.5ex \ULset}

\begin{document}

\title{Extracting strange quark freeze-out information in Pb+Pb collisions at $\sqrt{s_{NN}}$=2.76 TeV from $\phi$ and $\Omega$ production}
\author{Jie Pu}
\affiliation{School of Physics and Astronomy and
Shanghai Key Laboratory for Particle Physics and Cosmology,
Shanghai Jiao Tong University, Shanghai 200240, China}
\author{Kai-Jia Sun}
\affiliation{School of Physics and Astronomy and
Shanghai Key Laboratory for Particle Physics and Cosmology,
Shanghai Jiao Tong University, Shanghai 200240, China}
\author{Lie-Wen Chen\footnote{%
Corresponding author: lwchen$@$sjtu.edu.cn}}
\affiliation{School of Physics and Astronomy and
Shanghai Key Laboratory for Particle Physics and Cosmology,
Shanghai Jiao Tong University, Shanghai 200240, China}
\date{\today}

\begin{abstract}
Using a covariant quark coalescence model combined with a blast-wave-like
analytical parametrization for (anti-)strange quark phase-space freeze-out
configuration, we extract information on strange quark freeze-out dynamics
in Pb+Pb collisions at $\sqrt{s_{NN}}$=2.76 TeV by fitting the measured
transverse momentum spectra and elliptic flows~($v_2$) of $\phi$
mesons and $\Omega$ baryons.
We find that although both the measured and calculated $v_2$ of $\phi$ and
$\Omega$ satisfy the number-of-constituent-quark (NCQ) scaling, the NCQ-scaled
$v_2$ is significantly smaller than the $v_2$ of strange quarks, implying
that the NCQ-scaled $v_2$ of $\phi$ and $\Omega$ cannot be simply
identified as the $v_2$ of strange quarks at hadronization.
Meanwhile, our results indicate that the covariant quark coalescence model
can nicely describe the spectra and elliptic flows of $\phi$ and $\Omega$
simultaneously, suggesting the coalescence mechanism is still valid for
$\phi$ and $\Omega$ production in Pb+Pb collisions at LHC energies.
\end{abstract}

\maketitle

\section{Introduction}
\label{introduction}
The main goal of relativistic heavy-ion collisions, including those
being carried out at Relativistic Heavy-Ion Collider (RHIC) and Large Hadron
Collider (LHC), is to explore the Quantum Chromodynamics (QCD) phase diagram,
especially the properties of deconfined quark-gluon plasma (QGP) that could be
created in these collisions and its transition to hadronic
matter~\cite{Jac12,Shu17}.
Results of {\it ab initio} lattice QCD (LQCD) simulations~\cite{Aok06,Din15,Baz12,Baz14,Baz17}
and effective model approaches~\cite{Fuk11,Mun16} have provided important
insights on the QCD phase diagram.
Experimentally, however, the QGP cannot be probed directly since partons are
confined to form hadrons via hadronization during the dynamical evolution of heavy-ion
collisions.
Therefore, it is particularly important to study
the production of some special particles which have small final hadronic interactions
and thus could carry important information on the early QGP dynamics in relativistic
heavy-ion collisions.
The multistrange hadrons, e.g., the $\phi $ meson which carries
hidden strangeness ($s\bar s$) and the $\Omega$ baryon which consists of three
valence strange quarks ($sss$, i.e., $\Omega^-$) or anti-strange quarks
($\bar s \bar s \bar s$, i.e., $\bar \Omega ^+$), are
such particles, because they have small hadronic interaction
cross sections and are little affected by re-scattering effects in later hadronic
stage of the collisions~\cite{Sho85,Sin86,Hec98,Bas99,Che03,Bia81,Mul72}.
Furthermore, since both $\phi$ meson and $\Omega$ baryon consist solely of
(anti-)strange quarks, their production thus provides an ideal probe to extract
the strange quark freeze-out information at hadronization
in relativistic heavy-ion collisions.

The masses of strange quarks are comparable to the temperature of the
QGP and they are thus expected to be abundantly produced from quark and gluon
inelastic scattering in the QGP, and the strangeness enhancement is thus proposed
as one of the signatures for the QGP formation in relativistic heavy-ion
collisions~\cite{Raf82,Koc86}. In the past decades, strangeness production in
relativistic heavy-ion collisions has been a topic of great interest and
significant progress has been made in understanding the strangeness dynamics
and the QGP properties (see, e.g., Refs.~\cite{Koc17,Ko18,Blu18} for recent review).

Compared to the yield and invariant transverse
momentum spectrum, the elliptic flow ($v_2$), which is the second Fourier
coefficient of the azimuthal distribution of the emitted
particles~\cite{Vol96,Pos98}, is more sensitive to the early stage dynamics of
heavy-ion collisions~\cite{Oll92,Sor97,Dan98,Zha99}.
Of particular interest is that the observed elliptic flows of identified hadrons
in heavy-ion collisions at RHIC and LHC were found to satisfy the
number-of-constituent-quark (NCQ) scaling; that is, the elliptic flow per quark
is the same at the same transverse momentum per quark.
As shown in Refs.~\cite{Gre03,Fri03,Mol03,Fri08},
such a scaling of hadron elliptic flows according to their constituent quark
numbers can be understood via a unique hadronization mechanism, i.e., the
quark recombination/coalescence. The quark coalescence mechanism is also supported
by the observed anomalously large enhancement of baryon to meson ratio at
intermediate transverse momenta~\cite{Gre03,Fri03,Mol03,Fri08,Hwa03} as well
as the scaling relations observed among higher-order hadron anisotropic
flows~\cite{STAR03,STAR05,Che04,Kol04}. These findings provide a strong indication
that the quark degrees of freedom are dominant at the time of hadronization and
the partonic collectivity has been developed during the partonic evolution
prior to hadronization.

The production of $\phi$ mesons and $\Omega$ baryons in relativistic heavy-ion
collisions has been extensively investigated in the past decades~\cite{Hec98,Che06,%
Ada04-OmgXi,Ada05-OmgXiv2,Alt05-NA49Omg,Abe07-phiV2,Abe14-OmgXi,Abe15-phi,ALICE15v2,%
Ada16-OmgPhiBES,Ada16-OmgPhiV2,ChenJH06,ChenJH08,Hua09,He10,Cho17,Ye17,Jin18},
and this has significantly deepened our understanding on the strangeness
dynamics and the QGP properties.
Although the elliptic flow of parton degrees of freedom cannot be directly measured
experimentally, the NCQ-scaled elliptic flow of hadrons is believed to reflect
that of the constituent quarks at hadronization.
In fact, according to the naive momentum-space quark coalescence
model~\cite{Mol03,Che04,Kol04} in which only the quarks with equal momentum are
allowed to coalesce, the obtained NCQ-scaled $v_2$ of hadrons should be equal to
$v_2$ of the constituent quarks.
On the other hand, a more realistic dynamical quark coalescence model~\cite{Che06}
which is based on the quark phase-space information from a multiphase transport
(AMPT) model~\cite{Lin05} has been used to study the production and anisotropic flows of $\phi$
and $\Omega$ in Au+Au collisions at RHIC energies, and it is found that the NCQ-scaled
$v_2$ of $\phi$ and $\Omega$ are significantly smaller than that of strange quarks.
However, that work is failed to describe the transverse momentum spectra of
$\phi$ and $\Omega$~\cite{Che06}.

In addition, although the $v_2$ of $\phi$ mesons nicely follows the constituent
quark number scaling in Au+Au collisions at RHIC energies, the scaling tends
to be violated in Pb+Pb collisions at LHC energies based on the measured $v_2$ of
$\phi $ mesons and protons~\cite{ALICE15v2}, and this causes the discussion about
if the quark coalescence as a relevant particle production mechanism is still
valid or not in heavy-ion collisions at LHC energies~\cite{Cho17}.
Therefore, it is interesting to see if both the spectra and elliptic flows of
$\phi$ and $\Omega$ in Pb+Pb collisions at LHC energies can be simultaneously
described within a more realistic quark coalescence model, and thus to explore
the possibility of quantitatively extracting the strange quark freeze-out
information at hadronization from the measured data of $\phi$ and $\Omega$.
This is the main motivation of the present work.

In this work, we extend the covariant coalescence model~\cite{Dov91} combined
with a blast-wave-like~\cite{Ret04} analytical parametrization for constituent particle
phase-space freeze-out configuration,
which has been successfully applied recently to describe the (anti-)light-(hyper)nuclei
production in relativistic heavy-ion collisions via (anti-)nucleon (and/or hyperon)
coalescence~\cite{Sun15-Li5,Sun16-LamH4,Sun16-diLam},
to describe the transverse momentum spectra and elliptic flows of
$\phi $ and $\Omega$ in Pb+Pb collisions at $\sqrt{s_{NN}}$=2.76 TeV via quark
coalescence.
Our results indicate that the quark coalescence model can nicely describe
both the spectra and elliptic flows of $\phi$ and $\Omega$ simultaneously,
suggesting the quark coalescence mechanism is still valid for $\phi$ and
$\Omega$ production in Pb+Pb collisions at LHC energies.
We also find that both the measured and calculated $v_2$ of $\phi$ and
$\Omega$ satisfy the NCQ scaling, but the NCQ-scaled $v_2$ is significantly
smaller than the $v_2$ of strange quarks by a factor of about $1.35$ in
centrality $10-20\%$ Pb+Pb collisions at $\sqrt{s_{NN}} = 2.76$ TeV.
Furthermore, the strange quark freeze-out information is obtained.

The paper is organized as follows:
In Sec.~\ref{Sec:Model}, we introduce the covariant coalescence model combined with
a blast-wave-like analytical parametrization for (anti-)strange
quark phase-space freeze-out configuration.
We then apply the model to describe the transverse momentum spectra and
elliptic flows of $\phi$ and $\Omega^-$ in centrality $10-20\%$ Pb+Pb collisions at
$\sqrt{s_{NN}} = 2.76$ TeV, and then the obtained results are presented and discussed
in Sec.~\ref{Sec:Result}. Finally, we summarize our conclusions in Sec~\ref{Sec:Summary}.

\section{Model and method}
\label{Sec:Model}

In this work, the covariant coalescence model~\cite{Dov91} combined with
a blast-wave-like analytical parametrization~\cite{Ret04} for (anti-)strange
quark phase-space freeze-out configuration is used to describe
the production of $\phi$ and $\Omega$ in relativistic heavy-ion collisions.
In particular, for $\phi$ and $\Omega$ production at mid-rapidity in Pb+Pb
collisions at $\sqrt{s_{NN}}=2.76$~TeV considered here,
we assume a longitudinal boost-invariant expansion for the (anti-)strange quarks
and the Lorentz invariant one-particle momentum distribution is then given by
\begin{eqnarray}
E\frac{d^3N}{d^3p}=\frac{d^3N}{p_Tdp_T d\phi_p dy } =  \int d^4x S(x,p),
\end{eqnarray}
where $S(x,p)$ is the emission function and it is taken to be a blast-wave-like
parametrization as~\cite{Ret04}
\begin{eqnarray}
S(x,p)d^4x = m_T cosh(\eta-y)f(x,p)J(\tau)d\tau d\eta rdrd\phi_s.
\end{eqnarray}
In above expressions, we use longitudinal proper time $\tau = \sqrt{t^2-z^2}$,
spacetime rapidity $\eta = \frac{1}{2} \text{ln}\frac{t-z}{t+z}$, cylindrical
coordinates ($r$, $\phi_s$), rapidity $y=\frac{1}{2}\ln (\frac{E+p_z}{E-p_z})$,
transverse momentum ($p_T,\phi_p$), and transverse mass $m_T=\sqrt{m^2+p_T^2}$.
The statistical distribution function $f(x,p)$ is given by
\begin{eqnarray}
f(x,p)=g(2\pi)^{-3}[\exp(p^{\mu}u_{\mu}/kT)/\xi \pm 1]^{-1},
\end{eqnarray}
where $g$ is statistical degeneracy factor including spin and color degrees
of freedom, $p^\mu$ is the four-momentum of the
emitted particle, $u_{\mu}$ is the four-velocity of a fluid element in the
fireball, $T$ is the local temperature and $\xi$ is the fugacity.
The $p^{\mu}u_{\mu}$ is the energy in the local rest frame of the fluid and
reads
\begin{eqnarray}
p^{\mu}u_{\mu}=m_T\cosh\rho\cosh(\eta-y)-p_T\sinh\rho\cos(\phi_p-\phi_b),
\end{eqnarray}
where $\rho$ is the transverse rapidity distribution
(transverse flow profile) of the fluid element in the fireball,
$\phi_p$ is azimuthal direction of the emitted particle, and $\phi_b$ is
azimuthal direction of the transverse flow which is different from the
spatial azimuthal angle $\phi_s$. We also assume the freeze-out proper
time follows a Gaussian distribution~\cite{Ret04}
\begin{eqnarray}
J(\tau)=\frac{1}{\Delta \tau \sqrt{2\pi}}\exp\bigg(-\frac{(\tau-\tau_0)^2}{2(\Delta \tau)^2}\bigg)
\end{eqnarray}
with a mean value $\tau_0$ and a dispersion $\Delta \tau$.
More detailed information can be found in Refs.~\cite{Sun15-Li5,Sun16-LamH4,Sun16-diLam}.

In order to improve the description on the measured $\phi $ and $\Omega $ elliptic flows
at higher transverse momenta ($p_T\gtrsim 2$ GeV/c), an $r$-dependent coefficient
$c_1\exp(-r^2/c_2^2)$ is introduced in the transverse rapidity flow profile, and
the flow profile is parameterized as
\begin{eqnarray}
\rho=\rho_0 \tilde{r}\bigg[1+c_1e^{-\frac{r^2}{c_2^2}}\cos(2\phi_b)\bigg],
\label{Eq:rho}
\end{eqnarray}
where $\rho_0$ is the isotropic part of the transverse rapidity flow,
$c_1$ is introduced to describe anisotropy of transverse flow, $c_2$
denotes a suppression of anisotropy at larger $r$ where the
(anti-)strange quarks in the local cell have a larger averaged value
of $p_T$ due to the larger transverse velocity of the cell in the fireball,
and $ \tilde{r}$ is the ``normalized elliptical radius''~\cite{Ret04}
\begin{eqnarray}
\tilde{r}=\sqrt{\frac{[r\cos\phi_s]^2}{R_x^2}+\frac{[r\sin\phi_s]^2}{R_y^2}},
\end{eqnarray}
where $R_x=R_0(1+s_2)$ is the minor axis of the ellipse, $R_y=R_0(1-s_2)$
is the major axis, and $s_2$ denotes the geometric anisotropy. $R_0$ is
the transverse radius of the fireball.
The phase-space freeze-out
configuration of the constituent particles are thus determined by nine parameters:
$T$, $\rho_0$, $R_0$, $\tau_0$, $\Delta \tau$, $\xi$, $c_1$, $c_2$, $s_2$.

The Lorentz invariant one-particle momentum distribution can be
decomposed as~\cite{Vol96,Pos98}
\begin{eqnarray}
\begin{split}
E\frac{\mathrm{d}^3N}{\mathrm{d}^3p}=&\frac{\mathrm{d}^3N}{p_T\mathrm{d}p_T \mathrm{d}\phi_p \mathrm{d}y }\\
=&\frac{1}{2\pi}\frac{\mathrm{d}^2N}{p_T\mathrm{d}p_T \mathrm{d}y }\left[1+\sum_{n=1}^\infty 2v_n(p_T,y)\cos(n\phi_p)\right],\\
\end{split}
\end{eqnarray}
where $v_n$ denotes the anisotropic flows:
\begin{eqnarray}
\begin{split}
v_n=\langle \cos(n\phi_p) \rangle=\frac{\int\frac{\mathrm{d}^3N}{\mathrm{d}^2 p_T \mathrm{d}y}\cos(n\phi_p)\mathrm{d}\phi_p}{\int\frac{\mathrm{d}^3N}{\mathrm{d}^2 p_T\mathrm{d}y}\mathrm{d}\phi_p}.
\end{split}
\label{Eq:flow}
\end{eqnarray}
The famous elliptic flow~($v_2$) corresponds to the second Fourier coefficient
of the azimuthal distribution of the emitted particles.

With the above phase-space freeze-out information for constituent particles,
we can use the covariant coalescence model to calculate the invariant momentum
distribution of clusters. In the coalescence model, the probability for
producing a cluster is determined by the overlap of its Wigner phase-space
density (Wigner function) with the constituent particle phase-space
distribution at freeze-out. If $M$ particles are coalesced into a cluster,
the invariant differential transverse momentum distribution of the cluster
can be obtained as
\begin{eqnarray}
E\frac{\mathrm{d}^3N_c}{\mathrm{d}^3P}&=&Eg_c\int  \bigg(\prod_{i=1}^{M} \frac{\mathrm{d}^3p_i }{E_i}\mathrm{d}^4x_iS(x_i,p_i)\bigg)\times \notag \\
&&\rho_c^W(x_1,...,x_M;p_1,...,p_M)\delta^3\bigg(\mathbf{P}-\sum_{i=1}^M\mathbf{p_i}\bigg),  \notag \\
\label{Eq:Coal}
\end{eqnarray}
where $N_c$ is the cluster multiplicity, $E$ ($\mathbf{P}$) is its
energy (momentum), $\delta$-function is adopted to ensure momentum conservation,
$g_c$ is the coalescence factor including the spin and color degrees of
freedom and is is expressed as $g_c=\frac{2J+1}{2^M3^M}$~\cite{Sat81},
and $\rho_c^W$ is the Wigner function of the
cluster. In this work, the harmonic oscillator wave functions are assumed
for the cluster and its Wigner function is
\begin{eqnarray}
\begin{split}
&\rho_c^W\left(x_1,\cdot \cdot \cdot,x_M;p_1,\cdot \cdot \cdot,p_M\right)\\
=&\rho ^{W}(q_{1},\cdot \cdot \cdot ,q_{M-1},k_1,\cdot
\cdot \cdot ,k_{M-1})\\
=&8^{M-1}\exp \bigg[-\sum_{i=1}^{M-1}(q_{i}^{2}/\sigma _{i}^{2}+\sigma_{i}^{2}k_i^{2})\bigg],\\
\end{split}
\end{eqnarray}
where $\sigma_i^2 = (\mu_{i} w)^{-1}$, $\mu_{i-1}= \frac{i}{i-1} \frac{m_i\sum_{k=1}^{i-1}m_k}{\sum_{k=1}^{k=i}m_k}$ $(i\geq2)$
is the reduced mass in the center-of-mass frame,
$q_i=\sqrt{\frac{i}{i+1}}(\frac{\sum_{k=1}^{i}m_kx_k}{\sum_{k=1}^{i}m_k}-x_{i+1})$
is the relative coordinate, $k_i$ is the relative momentum, and $w$ is the harmonic
oscillator frequency which is related to the root-mean-square radius
of the cluster as follows \cite{Sun15-Li5}
\begin{eqnarray}
\langle r^2_M \rangle = \frac{3}{2M w}\bigg[\sum_{i=1}^M \frac{1}{m_i} -\frac{M}{\sum_{i=1}^M m_i}\bigg].
\end{eqnarray}
The integral~(\ref{Eq:Coal}) can be directly calculated through multi-dimensional
numerical integration by Monte-Carlo method \cite{Sun15-Li5,Lep78}. The cluster elliptic
flow can be calculated from Eq.~(\ref{Eq:flow}) and Eq.~(\ref{Eq:Coal}).
It should be emphasized that since the constituent particles may have different
freeze-out time, in the numerical calculation, the  constituent particles that freeze
out earlier are allowed to propagate freely until the time when the last constituent
particle in the cluster freezes out in order to make the coalescence
at equal time in the rest frame of the cluster~\cite{Mat97,Che06,Sun15-Li5}.

\section{result and discussion}
\label{Sec:Result}

By using the model and method introduced above, we can extract the phase-space
freeze-out information, namely, $T$, $\rho_0$, $R_0$, $\tau_0$, $\Delta \tau$,
$\xi$, $c_1$, $c_2$ and $s_2$, of (anti-)strange quarks in
Pb+Pb collisions at $\sqrt{s_{NN}}=2.76$~TeV by fitting the experimental
data on $\phi$ and $\Omega$ production.
In the following,
the mass of (anti-)strange quark is taken to be $500$ MeV, and the root-mean-square
radii of $\phi$ and $\Omega$ are taken to be $0.87$ fm and $1.0$ fm~\cite{Sun17-COAL-SH},
respectively.
The coalescence factors $g_c$ including spin and color degrees of
freedom are $3/(3^2\times2^2)$ and $4/(3^3\times2^3)$ for $\phi$
and $\Omega$, respectively.
The details can be found in Ref.~\cite{Sun17-COAL-SH}.
In addition, in the present work, the anti-strange quarks are assumed to have the
same freeze-out parameters as strange quarks because the $\bar \Omega ^+ / \Omega ^-$
ratio is close to unity in Pb+Pb collisions at $\sqrt{s_{NN}}=2.76$ TeV~\cite{Abe14-OmgXi}.

\begin{table}[tbp]
\caption{Parameters of the blast-wave-like analytical parametrization for
mid-rapidity (anti-)strange quark phase-space freeze-out configurations in
centrality $10-20\%$ Pb+Pb collisions at $\protect\sqrt{s_{NN}}=2.76$ TeV. }
\begin{tabular}{ccccccccccc}
        \hline \hline
          & T(MeV) & $\rho_0$ & $R_0$(fm) & $\tau_0$(fm/c) &  $\Delta \tau$(fm/c)   & $\xi_s$   \\
         \hline
        FOPb-s  & 154  & 1.06  & 14.8  & 13.0  & 1.3  & 0.78  \\
        \hline
        &  $c_1 $  &  $ c_2$ (fm)  &  $ s_2 $ \\
        \hline
        FOPb-s  & 0.38 & 8.7  &-0.05 \\
        \hline  \hline
\end{tabular}
\label{ParamHadron}
\end{table}

By fitting the transverse momentum spectra and elliptic flows of $\phi$ and
$\Omega^-$~\cite{Abe14-OmgXi,Abe15-phi,ALICE15v2} in centrality $10-20\%$ Pb+Pb
collisions at $\protect\sqrt{s_{NN} }=2.76$ TeV simultaneously, the parameters
of (anti-)strange quark phase-space freeze-out configurations are extracted and
summarized as FOPb-s (\textbf{F}reeze-\textbf{O}ut in \textbf{Pb}+Pb collisions for \textbf{s}trange quarks)
shown in Table~\ref{ParamHadron}.
Here the local temperature is fixed as $T = 154$ MeV following the QCD transition
temperature obtained from the high-precision studies of the chiral and deconfinement
aspects of the QCD transition at zero baryon chemical potential~\cite{Baz12}, and
the extracted transverse flow parameter is $\rho_0=1.06$, transverse radius
is $R_0=14.8$ fm, the longitudinal proper time is $\tau_0=13.0$ fm/c, the time
dispersion is $\Delta \tau = 1.3$ fm/c, the fugacity of (anti-)strange quark is
$0.78$, the geometric anisotropy is $s_2=-0.05$ and the anisotropy parameters
are $c_1=0.38$ and  $c_2 =8.7$ fm.
These parameters give a quantitative description about the (anti-)strange
quark phase-space freeze-out configuration for centrality $10-20\%$ Pb+Pb collisions at
$\protect\sqrt{s_{NN} }=2.76$ TeV.

\begin{figure}[tbp]
\includegraphics[scale=0.45]{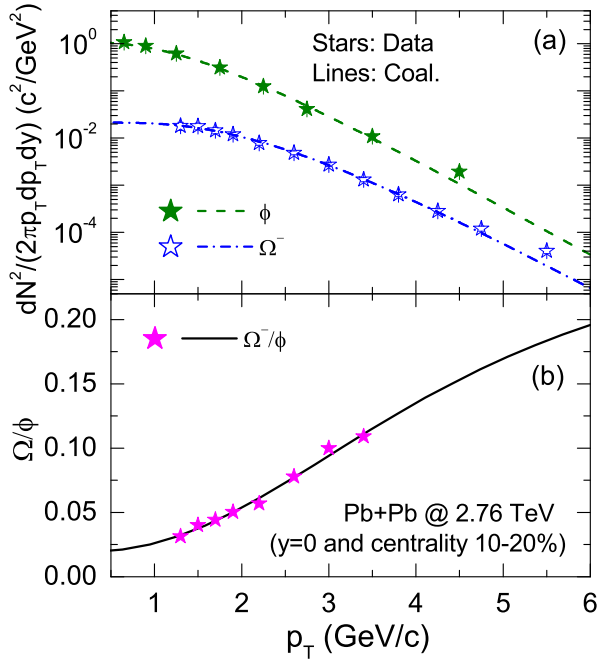}
\caption{(Color online) Transverse momentum distribution of mid-rapidity $\phi$ mesons
and $\Omega^-$ baryons in centrality $10-20\%$ Pb+Pb collisions
at $\sqrt{s_{NN}}$=2.76 TeV (a) and the corresponding yield ratio of $\Omega^-$ baryons
to $\phi$ mesons (b). The lines are from the quark coalescence model predictions and
the stars are experimental data taken from ALICE measurement~\cite{Abe14-OmgXi,Abe15-phi}.}
\label{spectrum}
\end{figure}

\begin{figure}[tbp]
\includegraphics[scale=0.31]{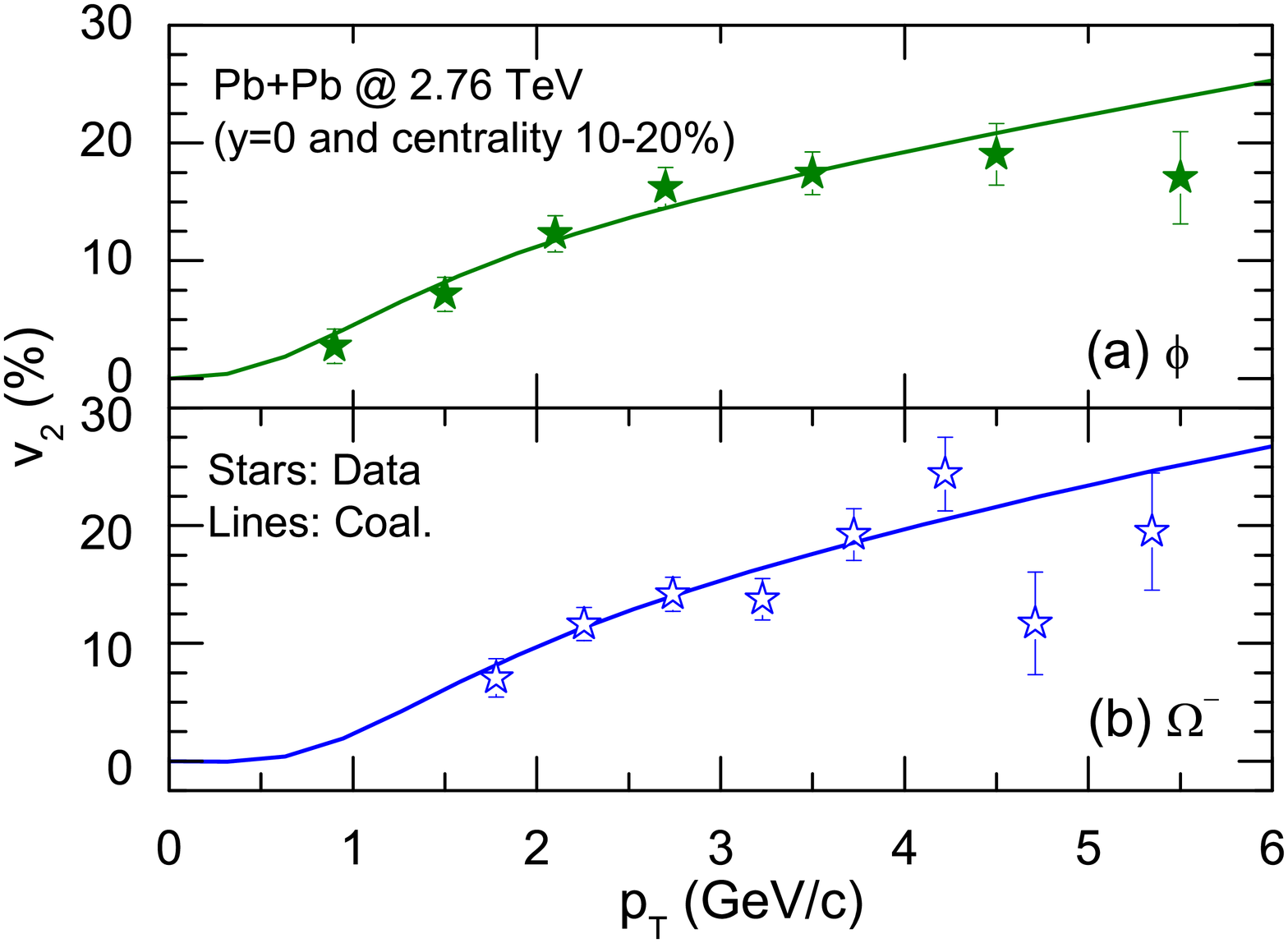}
\caption{(Color online) The measured (stars) and calculated (lines)
transverse momentum dependence of the elliptic flows ($v_2$) for mid-rapidity
$\phi$ mesons (a) and $\Omega^-$ baryons (b) in centrality $10-20\%$ Pb+Pb collisions
at $\sqrt{s_{NN}}$=2.76 TeV. The experimental results (stars) are taken from
ALICE measurement~\cite{ALICE15v2}.}
\label{flow}
\end{figure}

Shown in Fig.~\ref{spectrum} (a) are the experimental data and
theoretical calculations for the transverse momentum spectra of $\phi$ and
$\Omega^-$ in centrality $10-20\%$ Pb+Pb collisions at
$\sqrt{s_{NN}} = 2.76$ TeV. The corresponding results on the yield ratio
of $\Omega^-$ baryons to $\phi$ mesons as a function of transverse momentum are
shown in Fig.~\ref{spectrum} (b).
The experimental data are taken from ALICE measurement~\cite{Abe14-OmgXi,Abe15-phi}.
It is seen that
the present coalescence model predictions are in good agreement
with the experimental data from ALICE measurements.
In particular, the $\Omega^- / \phi$ ratio enhances with the transverse momentum,
e.g., its value changes from about $0.02$ at $p_T = 0.5$ GeV/c to about $0.1$
at $p_T = 3$ GeV/c, with an enhancement factor of about $5$.
This enhancement can be understood as a result of quark coalescence
mechanism~\cite{Gre03,Fri03,Mol03,Fri08,Hwa03} as in the case of observed anomalously
large anti-proton to pion ratio in central and mid-peripheral (centrality of about
$30\%$) Au+Au collisions at $\sqrt{s_{NN}}=200$ GeV~\cite{ppiPhenix}.
It should be pointed out that the enhancement can be also explained by the mass
effect (the $\Omega$ is heavier than the $\phi $) via fitting the data
with a Boltzmann-Gibbs blast-wave function~\cite{Abe15-phi,Sch93}.

Figure~\ref{flow} shows the experimental data and theoretical calculations on
the transverse momentum dependence of the elliptic flows of $\phi$ and
$\Omega^-$ in centrality $10-20\%$ Pb+Pb collisions at
$\sqrt{s_{NN}}$=2.76 TeV. The experimental data are taken from
the ALICE measurement~\cite{ALICE15v2}. One sees that the present quark
coalescence model can nicely describe the experimental data.
From Fig.~\ref{spectrum} and Fig.~\ref{flow},
it is seen that both the transverse momentum spectra and
elliptic flows of $\phi$ and $\Omega^-$ can be described very well in
the present covariant quark coalescence model with the same parameter set
FOPb-s, which provides important information on the phase-space freeze-out
configuration of mid-rapidity (anti-)strange quarks in centrality $10-20\%$
Pb+Pb collisions at $\sqrt{s_{NN}}$=2.76 TeV.
Our results demonstrate that
the quark coalescence mechanism is still valid for $\phi$ and
$\Omega$ production in Pb+Pb collisions at LHC energies.
The observed violation of the NCQ scaling for the $v_2$ of $\phi $ mesons
and protons in Pb+Pb collisions at LHC energies~\cite{ALICE15v2} probably
is due to the different final hadronic interactions for $\phi $ mesons
and protons in heavy-ion collisions at LHC energies.

\begin{figure}[tbp]
\includegraphics[scale=0.35]{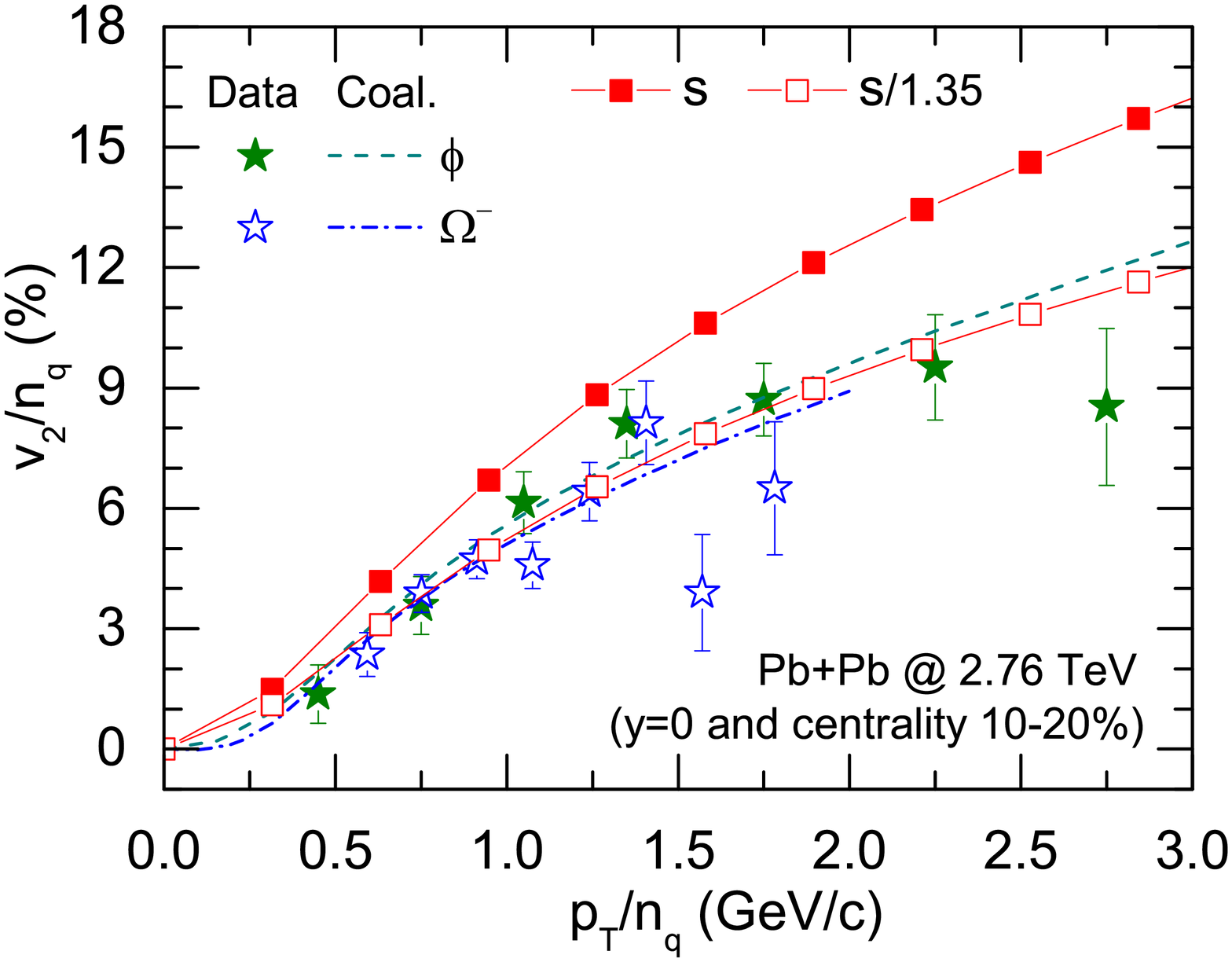}
\caption{(Color online) NCQ-scaled elliptic flow ($v_2/n_q$) as a function of
scaled transverse momentum ($p_T/n_q$) for mid-rapidity $\phi$ mesons and
$\Omega^-$ baryons in centrality $10-20\%$ Pb+Pb collisions at $\sqrt{s_{NN}}$=2.76 TeV.
Also shown are the corresponding results from quark coalescence model predictions
(dashed and dash-dotted lines) as well as the results for (anti-)strange quarks at
freeze-out (solid squares) and their scaled values (divided by a factor of $1.35$)
(open squares). The experimental data (stars) are taken from ALICE measurement~\cite{ALICE15v2}.}
\label{scal}
\end{figure}

Figure~\ref{scal} shows the measured and calculated NCQ-scaled elliptic flow
$v_2/n_q$ as a function of scaled transverse momentum $p_T/n_q$ for mid-rapidity
$\phi$ and $\Omega$ in centrality $10-20\%$ Pb+Pb collisions at
$\sqrt{s_{NN}}$=2.76 TeV. Also included in Fig.~\ref{scal} are the corresponding
results for (anti-)strange quarks at freeze-out as well as their scaled values
(divided by a factor of $1.35$).
It is seen that the measured and calculated elliptic flows of $\phi$ and $\Omega$
approximately satisfy the famous NCQ scaling relation. However, the NCQ-scaled
elliptic flows of $\phi$ and $\Omega$ are significantly smaller than that of the
coalescing (anti-)strange quarks. More quantitatively, we find that the elliptic
flow of (anti-)strange quarks is about $1.35$ times the NCQ-scaled elliptic
flows of $\phi$ and $\Omega$.
This is different from the prediction of the naive momentum-space quark coalescence
model~\cite{Mol03,Che04,Kol04} in which only the quarks with equal momentum are
allowed to coalesce. In this simple model, the momentum spectrum of $\phi$ meson
is proportional to the product of the momentum spectra of strange and anti-strange quarks,
leading to the result that the NCQ-scaled elliptic flow $v_2/n_q$ of $\phi$ mesons
equals to the $v_2$ of (anti-)strange quarks, which can be demonstrated from a simple
Fourier analysis~\cite{Mol03,Che04,Kol04}. The same conclusion is also obtained in the
case of $\Omega$ baryons. Therefore, in the naive momentum-space quark coalescence
model, the obtained NCQ-scaled $v_2$ of $\phi$ and $\Omega$
should be equal to $v_2$ of (anti-)strange quarks.

Unlike the naive momentum-space quark coalescence
model, in the present covariant quark coalescence model,
the effects of finite sizes of hadrons and nonzero relative momenta
of partons inside the hadrons have been encoded in the hadron Wigner
function in full phase-space and thus quarks with unequal momenta
can be coalesced into hadrons ($\phi$ and $\Omega$), and this may
smear the azimuthal distribution of the formed hadrons and thus leads
to a suppression of hadron elliptic flows.
The magnitude of the suppression is directly related to the internal
structure and size of the hadrons.
Although this phenomenon has actually been observed in previous
work~\cite{Che06} within a dynamical quark coalescence model using the parton
freeze-out information based on the AMPT transport model calculations,
the transverse momentum spectra of $\phi$ and $\Omega$ were failed
to be reproduced there.
Our results indicate that the NCQ-scaled elliptic flows of $\phi$ and $\Omega$
significantly underestimate the elliptic flow of (anti-)strange quarks and
they cannot be simply identified as the elliptic flow of the (anti-)strange
quarks in relativistic heavy-ion collisions.

\section{Conclusion}
\label{Sec:Summary}

Based on the covariant quark coalescence model with a blast-wave-like analytical
parametrization for the phase-space configuration of mid-rapidity (anti-)strange
quarks at freeze-out, we have extracted information of strange quark freeze-out
dynamics in centrality $10-20\%$ Pb+Pb collisions at $\sqrt{s_{NN}}$=2.76 TeV by
simultaneously fitting the transverse momentum spectra and
elliptic flows of $\phi$ and $\Omega$.
We have found that our model can successfully describe the experimental data
on both the transverse momentum spectra and elliptic flows of $\phi$ and $\Omega$,
demonstrating that the quark coalescence mechanism is still
valid for $\phi$ and $\Omega$ production in heavy-ion collisions at LHC energies.

Our results indicate that the measured and calculated elliptic flows of $\phi$
and $\Omega$ approximately satisfy the famous NCQ scaling relation, but unlike
the prediction of the naive momentum-space quark coalescence model, the NCQ-scaled
elliptic flows of $\phi$ and $\Omega$ are significantly smaller than that of the
coalescing (anti-)strange quarks, with the latter being about
$1.35$ times the former.
This means that one cannot simply identify the experimentally measured NCQ-scaled
elliptic flows of $\phi$ and $\Omega$ as the elliptic flow of (anti-)strange
quarks in relativistic heavy-ion collisions.

The present work provides useful information on the strangeness freeze-out
dynamics in relativistic heavy-ion collisions.
The model and method in the present work
can be further applied to heavy-ion collisions at energies of beam energy
scan~(BES) program at STAR/RHIC. Such studies are in progress and will be
reported elsewhere.

\begin{acknowledgments}
The authors thank Professor Che Ming Ko for helpful discussions.
This work was supported in part by the National Natural Science
Foundation of China under Grant No. 11625521, the Major State Basic Research
Development Program (973 Program) in China under Contract No.
2015CB856904, the Program for Professor of Special Appointment (Eastern
Scholar) at Shanghai Institutions of Higher Learning, Key Laboratory
for Particle Physics, Astrophysics and Cosmology, Ministry of
Education, China, and the Science and Technology Commission of
Shanghai Municipality (11DZ2260700).
\end{acknowledgments}

\end{document}